# Optimization of Simulations and Activities for a New Introductory Quantum Mechanics Curriculum

Antje Kohnle, Charles Baily, Christopher Hooley and Bruce Torrance

*School of Physics and Astronomy, University of St Andrews, North Haugh, St Andrews, KY16 9SS, United Kingdom*

**Abstract:** The Institute of Physics New Quantum Curriculum (quantumphysics.iop.org) consists of online texts and interactive simulations with accompanying activities for an introductory course in quantum mechanics starting from two-level systems. Observation sessions and analysis of homework and survey responses from in-class trials were used to optimize the simulations and activities in terms of clarity, ease-of-use, promoting exploration, sense-making and linking of multiple representations. This work led to revisions of simulations and activities and general design principles which have been incorporated wherever applicable. This article describes the optimization of one of the simulation controls and the refinement of activities to help students make direct connections between multiple representations.

## INTRODUCTION

The Institute of Physics New Quantum Curriculum (quantumphysics.iop.org) consists of online resources for the learning and teaching of introductory quantum mechanics starting from two-level systems. This approach immediately immerses students in the concepts of quantum mechanics by focusing on experiments that have no classical explanation. It allows from the start a discussion of the physical interpretation of quantum mechanics and recent developments such as quantum information theory. The text articles have been written by researchers in quantum information theory and foundations of quantum mechanics. One of us (AK) designed the interactive simulations and activities (17 in total) that are part of this resource, and which cover the topics of linear algebra, fundamental quantum mechanics concepts, single photon interference, the Bloch sphere representation, entanglement, local hidden variables and quantum information.

Computer simulations can promote engaged exploration and sense-making, and can help students make connections between multiple representations, including those not readily visible in the real world. [1] High levels of interactivity and direct feedback allow students to explore relationships between different quantities. Simulations can be particularly useful for the learning and teaching of quantum mechanics due to its counterintuitive results and its abstract nature far removed from everyday experience. Research-based interactive simulations for quantum mechanics have been developed and shown to improve student understanding. [2-4]

The New Quantum Curriculum simulations make use of principles of interface design from previous studies. [3, 5-7] Each simulation has two tabs which are used to toggle between two different views: The *Simulation* view contains introductory text and interactive controls; the *Step-by-step Exploration* view allows the user to step through detailed text explanations with animated highlighting. The combination of the two allows the simulations to be used as self-contained instructional resources. Figure 1 shows a screenshot of the "Entangled spin ½ particle pairs versus hidden variables" simulation (referred to as "hidden variable simulation" in what follows). In this simulation, students can send particle pairs through two Stern-Gerlach apparatuses to assess whether a simple hidden variable theory using instruction sets would agree with the measurement outcomes predicted by quantum theory. The activities were designed to promote guided exploration and sense-making. They promote initial free exploration by prompting students at the start to describe what they have discovered by simply playing with the various simulation controls.

Trialing simulations and activities with students at the appropriate level is key to tailoring these resources to meet students' needs. We have iteratively refined the simulations and activities using individual student observation sessions (for 16 simulations) and in-class trials (for 3 simulations), with the aim of optimizing the simulations in terms of ease-of-use, the clarity of physical setups, graphs, displayed quantities and text explanations, and promoting exploration and sense making. The activities were similarly refined for clarity and the promotion of guided exploration, while at the same time providing sufficient scaffolding for students to progress from simpler to more complex situations. In this article, we focus only on a small subset of these aims, namely the optimization of simulation controls to be intuitive (thus ensuring a greater focus on the actual content of the simulations), and changes to the activities that help students make direct connections between multiple representations.

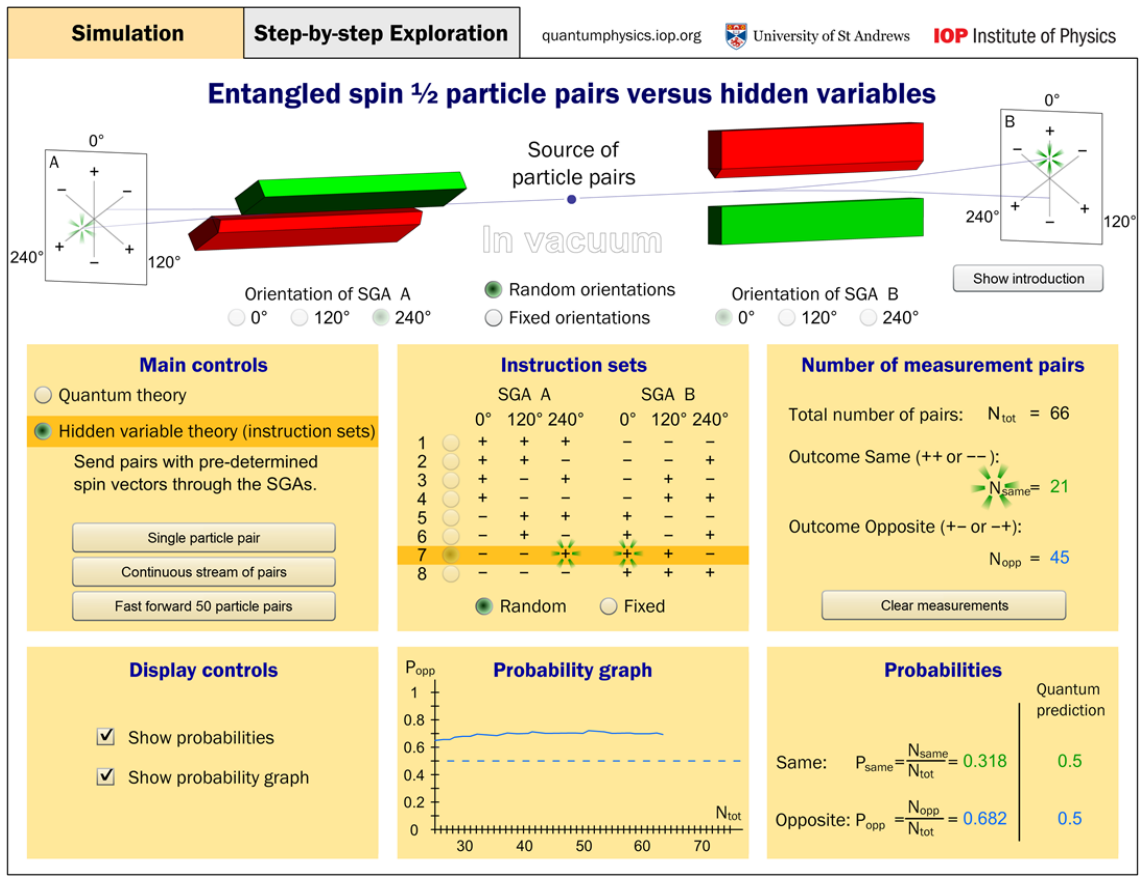

**FIGURE 1.** A screenshot of the "Entangled spin ½ particle pairs versus hidden variables" simulation. Students may toggle between the *Simulation* and *Step-by-step Exploration* tabs at the top, and collect experimental data in order to compare and contrast between a simple hidden-variable theory and the predictions of quantum mechanics.

## METHODOLOGY

We conducted observation sessions with student volunteers from the University of St Andrews Quantum Physics course. This is an introductory course roughly equivalent to a Modern Physics course for sophomores in the United States, and is likely to be the first university course in quantum mechanics taken by our students. Roughly a third of the sessions were carried out in February 2013 prior to the in-class trials (see below), the others in May 2013 after the end of the semester. Much of the simulation content was new to these student volunteers, all of whom were physics majors.

We conducted 19 two-hour observation sessions with a total of 17 student volunteers, which were recorded for both audio and screencapture. We were able to trial all of the simulations and activities except one (16 in total) in these sessions. Five simulations were tested by 1 student, five simulations by 2 students, one simulation by 3 students, two simulations by 4 students and three simulations by 5 students. Typically, two simulations were used in a single session, but in some cases one or three were used. For a number of the simulations and activities there was sufficient time between trials to implement minor changes based on our observations of student difficulties prior to testing them again with subsequent students.

In these sessions, students were asked to freely explore a simulation while thinking aloud and describing what they were investigating, and to explain what they understood or found confusing; they then worked on the activity associated with the simulation. When students struggled with understanding a particular aspect of the simulation or activity, we often asked clarifying questions to determine why, and asked how the simulation or activity (or the link between them) might be improved. We observed whether students took notice of all the relevant controls and how they interacted with them (e.g. if they expected controls to function differently than designed). We also observed whether and how students made use of the simulation to answer the questions in the activities, such as collecting data and then comparing the outcomes with their calculations, or by actually configuring the simulation to mirror the situation described in the problem statement.

Afterwards, students completed a follow-up survey asking them to rate the ease-of-use and the clarity of various aspects of the simulation and the clarity of the associated activity. The survey also prompted them to elaborate on which aspects they found confusing and to make further suggestions for improvement.

In addition to these sessions, three of the simulations were used during the Spring 2013 Quantum Physics course at St Andrews (with 94 students, almost all of whom are physics majors). The course content was revised to include parts of the New Quantum Curriculum. We used the hidden variable simulation (shown in Fig. 1) and the "Interferometer experiments with photons, particles and waves" simulation in computer classroom workshops, and the "Entanglement: the nature of quantum correlations" simulation as part of a homework assignment. Two of these simulations (interferometer experiments, hidden variables) were also used in homework assignments in a Modern Physics course at the University of Colorado Boulder (N=77) a few weeks prior to their use at St Andrews. Students at both institutions were asked to respond to the same post-interview survey questions used in the observation sessions. We then analyzed and grouped student responses to determine common difficulties with the problem statements, and which aspects of the simulations were confusing and in need of improvement.

## OUTCOMES

In this section we describe the optimization of one of the simulation controls and changes made to the activities in order to help students make direct connections between multiple representations. We also provide preliminary evidence that these revisions have improved the simulations and activities.

**Optimizing the simulation interface:** Poorly-designed controls can lead to user frustration and a focus on the control itself instead of the content. [1, 6] In the hidden variable simulation, we initially only had "Single particle pair" and "Continuous stream of pairs" controls for students to send particles through the experiment (see the top two buttons in the Main controls panel in Fig. 1). Student feedback suggested that the limitations of these controls led to frustration, and made it difficult to compare experimental values and theoretical predictions. From the survey questions on suggestions for improvement, there were six comments (out of 40 suggestions for improvement) from Boulder students that it took too long to collect proper statistics, and that they would like some way to speed up data collection. A similar issue was encountered with the interferometer experiments simulation, where students can send single photons, electromagnetic waves and particles through a Mach-Zehnder interferometer. For this simulation, there were an additional 16 comments (out of 42 suggestions for improvement) pertaining to speeding up the detections.

In response to this feedback, we incorporated the "Fast forward 50 particle pairs" button into the hidden variable simulation (shown in the Main controls panel in Fig. 1) prior to its use in the St Andrews course. This control adds 50 counts to the number of measurement pairs all at once, and updates the calculated probabilities and the probability graph accordingly. Thus, meaningful comparisons between experimental and theoretical probabilities can be obtained more quickly. None of the St Andrews students' comments (59 in total) on suggestions for improvement for the hidden variable simulation pertained to the speed of data collection, suggesting that the additional control resolved this issue. Given that this was perceived as a problem in both simulations used at Boulder, we have now included a fast forward control in all of the simulations where data are collected.

In the observation sessions, most students began exploring the controls from top to bottom (hence starting with single fire mode), which justifies the ordering of the layout shown in Fig. 1, since we would like students to first make sense of a few initial data points before fast forwarding to larger data sets. Some students mentioned that they appreciated the ability to take data at their preferred rate, and that this helped them to understand the connection between the experimental values and theoretical predictions. One student working with another simulation where a fast forward button had been incorporated remarked that it was "[n]ice that you can have both a single particle and then you change to the continuous thing and it was also nice that you can just fast forward it so that you can just hit [the button] many times and see what it does in the limiting case."

**Refining the activities:** By using multiple representations to illustrate phenomena, interactive simulations can help students develop visual mental models and encourage them to make connections between different representations. [5, 7] We have found that explicitly asking students within the activities to make detailed comparisons of their calculations with the simulation encouraged them to make direct connections between mathematical and visual representations.

The activity for the hidden variable simulation used in the St Andrews Quantum Physics course included two questions asking students to compare their calculations with the results shown in the simulation (e.g., "...calculate these two probabilities. Compare your results with those shown in the simulation." and a second similar question). We found that a substantial fraction of students (30% and 34% respectively, N=81) did not comment on having made this comparison despite calculating the probabilities, or responded only superficially (16% and 30% respectively). Examples of these superficial responses to the comparison questions above were "These are consistent with the simulation predictions." and "These all agree with the values tended toward in the simulation."

Comparing calculations with experimental situations and data in the simulations is a key aspect of helping students to make connections between multiple representations, and interpreting their results physically or graphically. Thus, we modified all activities wherever applicable to include separately numbered questions for these comparisons, using formulations such as "Explain how you can see these results graphically in the simulation" or "Describe how you can see these results in the simulation, including a description of the experimental setup..."

In subsequent observation sessions, we found that students did not skip over these questions, and that they gave detailed and explicit responses. For example, one student using the "Uncertainty of spin measurement outcomes" simulation responded to the question "Explain how you can see that the measurement uncertainty is zero in the simulation" by pointing out two of the displayed graphs and stating: "The uncertainty graph is at zero, [and] the outcome uncertainty box, or display, is zero.". Another student working with the "Spin 1 particles in successive Stern-Gerlach experiments" simulation responded to a prompt to explain how specific results could be seen in the simulation by making a detailed sketch of the Stern-Gerlach apparatus, and wrote "The output beam labeling allows you to read the probabilities".

## CONCLUSIONS / FUTURE STEPS

The findings described here suggest that explicit questions asking students to make direct comparisons between their calculations and results shown in the simulation can encourage students to make connections between multiple representations. We have also found that adding controls to the simulations that allow students to collect data at a rapid pace can help them to make meaningful comparisons between experimental and theoretical quantities.

The observation sessions so far have been limited to a relatively small number of students from a single institution, and only a small number of simulations have been tested in courses. We will be conducting further observation studies at multiple institutions in the coming year, and ultimately evaluating the pedagogical effectiveness of the simulations when incorporated into quantum mechanics courses that develop the theory using two-level systems. Further refinements to the simulations and activities will be made based on the outcomes of these evaluations. Another avenue of development will be to create multiple versions of the activities, so they can be used either by individual students or in small groups where students would work collaboratively to complete the tasks.

## ACKNOWLEDGMENTS

We thank Noah Finkelstein at the University of Colorado Boulder for trialing two simulations in the Spring 2013 Modern Physics course. We gratefully acknowledge all of the students who participated in this study. We thank the Institute of Physics for funding this project, and developing and maintaining the New Quantum Curriculum website.